\documentclass{ws-procs9x6}
\newcommand{\lsim}{\raisebox{-0.7ex}{$\stackrel{\textstyle <}{\sim}$ }}
\newcommand{\gsim}{\raisebox{-0.7ex}{$\stackrel{\textstyle >}{\sim}$ }}

\def\si{^1 \hskip -0.04in S _0}
\def\siii{^3 \hskip -0.04in S _1}
\def\diii{^3 \hskip -0.04in D _1}

\begin{document}

\title{Few-Body Lattice Calculations}

\author{Martin J. Savage}

\address{Department of Physics, University of Washington, 
Seattle, WA 98195-1560.}

\begin{abstract}
I will discuss the recent progress toward computing few-body observables using
numerical lattice techniques.
The focus is overwhelmingly on the latest results from lattice QCD
calculations.
I present preliminary results from a lattice calculation of the central and
tensor potentials between B-mesons in the heavy-quark limit.
\end{abstract}


\bodymatter

\section{Introduction}
\label{sec:intro}

Perhaps the greatest challenge facing those of us working in the area of strong
interaction physics is to be able to rigorously
compute the properties and interactions of nuclei.
The many decades of theoretical and experimental investigations in nuclear
physics have, in many instances, provided a very precise phenomenology of the
strong interactions in the non-perturbative regime.  However, at this point in
time we have little understanding of much of this phenomenology in terms of the underlying
theory of the strong interactions, Quantum Chromo Dynamics (QCD).

The ultimate goal is to be able to rigorously compute the properties and
interactions of nuclei from QCD.  This includes determining how the structure
of nuclei depend upon the fundamental constants of nature.
Any nuclear observable is essentially a function of only five constants,
the length scale of the strong interactions, $\Lambda_{\rm QCD}$ , the quark
masses, $m_u$, $m_d$, $m_s$, and the electromagnetic coupling, $\alpha_e$ (at
low energies the dependence upon the top, bottom and charm quarks masses is
encapsulated in $\Lambda_{\rm QCD}$).
Perhaps as important,
we would then  be in the position to reliably
compute quantities that cannot be accessed, either directly or indirectly,
by experiment.

The only way to rigorously compute strong-interaction quantities in the
nonperturbative regime is with lattice QCD.  One starts with the QCD Lagrange
density and performs a Monte-Carlo evaluation of Euclidean space Green
functions directly from the path integral.  
To perform such an evaluation, space-time is latticized and computations are
performed in a finite volume, at finite lattice spacing, and at this point in
time, with quark masses that are larger than the physical quark masses.
To compute any given quantity, contractions are performed in which the
valence quarks that propagate on any given gauge-field configuration are ``tied
together''.  For simple processes such as nucleon-nucleon scattering, such
contractions do not require significant computer time compared with lattice or propagator
generation.  However, as
one explores processes involving more hadrons, the number of contractions grows rapidly
(for a nucleus with atomic number $A$ and charge $Z$, the number of
contractions is $(A+Z)!(2A-Z)!$),
and a direct lattice QCD calculation of the properties of a large nucleus 
is quite impractical simply due to the computational time required.
The way to proceed is to establish a small number of effective theories, each
of which have well-defined expansion parameters and can be shown to be the most
general form consistent with the symmetries of QCD.
Each theory must provide a complete description of nuclei over some range of atomic
number.  Calculations in two ``adjacent'' theories are performed for a range of
atomic numbers for which both theories converge.  One then matches coefficients
in one EFT to the calculations in the other EFT or to the lattice, and thereby one can make an
indirect, but rigorous connection between QCD and nuclei.
It appears that four different matchings are required:
\begin{enumerate}
\item 
{\bf Lattice QCD}.
Lattice QCD calculations of the properties of the very
lightest nuclei will be possible at some point in 
the not so distant future~\cite{Savage:2005ma}.
Calculations for $A\le 4$ as a function of the light-quark masses, would
uniquely define the interactions between nucleons up to and including the four-body
operators.
Depending on the desired precision, one could
possibly imagine calculations up to $A\sim 8$.

The chiral potentials and interactions have been determined out to the order
where four-body interactions contribute~\cite{Epelbaum:2006eu}.
As with the EFT constructions in the meson-sector and single-nucleon sector,
the number of counterterms proliferates with increasing order in the expansion,
and at some order one looses rigorous predictive power without external input.
Lattice QCD will make model independent determinations of these counterterms.

\item
{\bf Exact Many-Body Methods}.
During the past decade one has seen remarkable progress in the calculation of
nuclear properties using Green Function Monte-Carlo (GFMC) with the
$AV_{18}$-potential (e.g. Ref.~\cite{Pieper:2004qw}) 
and also the No-Core Shell Model (NCSM) (e.g. Ref.~\cite{Forssen:2004dk})
using chiral potentials.
Starting with the chiral potentials, which are the most general interactions
between nucleons consistent with QCD, one would calculate the properties of
nuclei as a function of all the parameters in the chiral potentials 
out to some given order in the chiral expansion.  A comparison between such calculations
and lattice QCD calculations will determine these parameters to some level
of precision.  These parameters can then be used in the calculation of nuclear
properties up to atomic numbers $A\sim 20-30$.
The computer time for these many-body theories suffers from the same $\sim
(A!)^2$ blow-up that lattice QCD does, and for a sufficiently large nucleus,
such calculations become impractical.

Another recent development that shows exceptional promise is the latticization
of the chiral effective field 
theories~\cite{Muller:1998rc,Muller:1999cp,Lee:2004si,Borasoy:2005yc,Seki:2005ns}.
This should provide  a model-independent calculation of nuclear processes once
matched to lattice QCD calculations.
\item
{\bf Coupled Cluster Calculations}.
In order to move to larger nuclei, $A\lsim 100$ a technique that has shown
promise is to implement a coupled-clusters expansion (e.g. Ref.~\cite{Wloch:2005qq}).
One uses the same chiral potential that will have been matched to lattice QCD
calculations, and then performs a diagonalization of the nuclear Hamiltonian,
after truncating the cluster expansion, which itself contains arbitrary coefficients.
The results of these calculations will be matched to those of the NCSM or GFMC
for $A\sim 20-30$
to determine the arbitrary coefficients.
This method is unlikely to be practical for very large atomic numbers.
\item
{\bf Density Functional Theory (??) and Very Large Nuclei}
To complete the periodic table one needs to have an effective theory that is
valid for very large nuclei and nuclear matter.
A candidate that has received recent attention is Density Function Theory
(DFT) (e.g. Refs.~\cite{Furnstahl:2004xn,Schwenk:2004hm}).  
It remains to be seen if this is in fact a viable candidate.
There is reason to hope that this will be useful because 
there is clearly a
density expansion in large nuclei with a power-counting that is consistent with
the 
Naive Dimensional Analysis (NDA) of Georgi and Manohar~\cite{Manohar:1983md}.
The application of DFT to large nuclei is presently the
least rigorously developed component of this program.

The latticized chiral theory mentioned previously can also be applied to the
infinite nuclear matter problem.  This work is still in the very earliest stages
of exploration, but this looks promising~\cite{Lee:2004si}.
\end{enumerate}

\section{Lattice QCD Calculations of Single Nucleon Properties}

While the lattice QCD calculations have historically focused on the meson
sector, both the properties of single mesons and the interactions between two
mesons, primarily due to limitations in computational power, the last few years
has seen an increasing number of precise calculations of the properties of
nucleons at the available pion masses. I wish to show two 
that will be of interest to the participants of this meeting.

\subsection{The Matrix Element of the Axial Current}
\begin{figure}
\begin{center}
\psfig{file=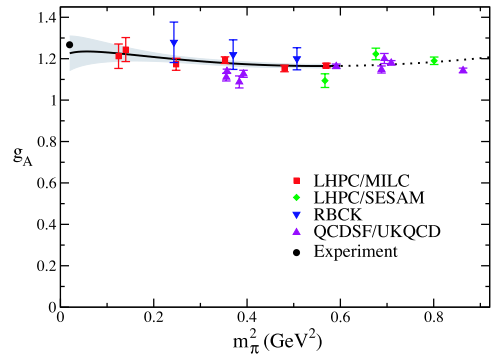,width=2.5in}
\end{center}
\caption{$g_A$ as a function of $m_\pi^2$.
This figure is reproduced with the permission of LHPC.
}
\label{fig:ga}
\end{figure}
The matrix element of the axial current in the nucleon, $g_A$,  is a fundamental
quantity in nuclear physics as it is related to the strength of the long-range
part of the nucleon-nucleon interaction via PCAC. 
During the last year LHPC~\cite{Edwards:2005ym} has for the first time calculated $g_A$ at small
enough pion masses where Heavy Baryon Chiral Perturbation
Theory (HB$\chi$PT) should converge.  The results of this calculation, previous
calculations and the physical value,
are shown in Fig.~\ref{fig:ga}. Also shown is the chiral
extrapolation along with is uncertainty (the shaded region).  
The physical value lies within the
range predicted by chiral extrapolation of the lattice calculations.

\subsection{The Neutron-Proton Mass Difference}

During the last year it was realized that one could use isospin-symmetric
lattices to compute isospin-breaking quantities to NLO in the chiral expansion~\cite{Beane:2006fk}.
This is achieved by performing partially-quenched (unphysical) calculations of the nucleon
mass~\cite{Beane:2002vq} in which the valence quark masses differ from the sea-quark masses
(those of the configurations) and determining counterterms in the
partially-quenched chiral Lagrangian.
The NPLQCD collaboration found that, in the absence of electromagnetic
interactions, the neutron-proton mass difference at the physical value of the
quark masses is  
\begin{eqnarray}
M_n-M_p \big|^{m_d-m_u} & = & +2.26 \pm 0.57\pm 0.42\pm 0.10~{\rm MeV}
\ \ \ ,
\end{eqnarray}
which is to be compared with estimates derived from the Cottingham sum-rule of 
$M_n-M_p \big|^{m_d-m_u}_{\rm physical} = +2.05 \pm 0.3$.
We see that the lattice determination is consistent with what is found in
nature.  The lattice calculation is expected to become significantly more
precise during the next year.

\section{Hadron Scattering from Lattice QCD}

To circumvent the Maiani-Testa theorem~\cite{Maiani:1990ca}, 
which states that one cannot compute Green functions
at infinite volume on the lattice and recover S-matrix elements except at
kinematic thresholds, one computes the energy-eigenstates of the two particle
system at finite volume to extract the scattering amplitude~\cite{Luscher:1986pf}.
The scattering amplitude of two pions in the $I=2$-channel has been the process
of choice to explore this technique, and to determine the reliability and
systematics of lattice calculations.

\subsection{$\pi\pi$ Scattering}

NPLQCD has calculated $\pi\pi$ scattering in the $I=2$ channel at relatively
low pion masses using the mixed-action technique of LHPC.  Domain-wall valence
propagators are calculated on MILC 
configurations containing staggered sea-quarks.
The lattice calculations
were performed with the {\it Chroma} software
suite~\cite{Edwards:2004sx,sse2} on the high-performance computing
systems at the Jefferson Laboratory (JLab).
\begin{figure}
\begin{center}
\psfig{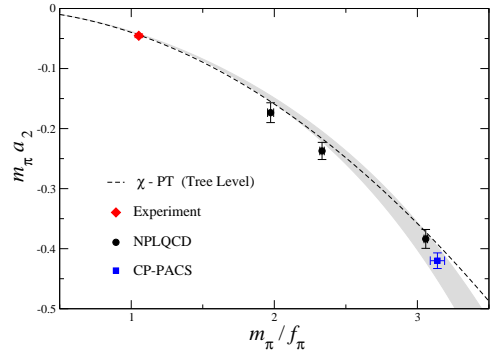}
\end{center}
\caption{The $\pi\pi$ scattering length in the $I=2$ channel as a function of
$m_\pi^2/f_\pi^2$.  
The dashed curve corresponds to the unique prediction of tree-level $\chi$PT,
while the shaded region is the fit to the results of the NPLQCD 
calculations~\protect\cite{Beane:2005rj}.
The NPLQCD calculations are performed at a single lattice spacing of $b\sim
0.125~{\rm fm}$.
}
\label{fig:pipi}
\end{figure}

This calculation demonstrates the predictive capabilities of lattice QCD
combined with low-energy EFT's. 
By writing the expansion of $m_\pi a_2$ as a function of $m_\pi^2/f_\pi^2$, 
it has been shown that  when inserting the values of $m_\pi$ and $f_\pi$ as
calculated on the lattice.
The lattice spacing effects in this mixed action calculation,
which naively appear at ${\cal O}(b^2)$, are further suppressed, appearing only
at higher orders~\cite{Chen:2005ab,Walker-Loud:2006sa}.  The first non-trivial
contributions from the partial-quenching in this calculation are shown to be
numerically very small, and therefore a clean extraction of the
counterterm in the $\chi$PT Lagrangian is possible.
The chiral extrapolation, as shown in Fig.~\ref{fig:pipi}, is found to have a
smaller uncertainty at the physical point than the experimental value.

\subsection{$K\pi$ Scattering }

Studying the low-energy interactions between kaons and pions with  $K^+\pi^-$
bound-states allows for an explicit exploration of the three-flavor structure of 
low-energy hadronic interactions, an aspect that is not directly probed 
in $\pi\pi$ scattering.
Experiments have been proposed by the DIRAC collaboration~\cite{DIRACprops}
to study $K\pi$ atoms
at CERN, J-PARC and GSI, the results of which would provide direct measurements or constraints
on combinations of the scattering lengths.
In the isospin limit, there are two isospin channels available to the $K\pi$
system, $I={1\over 2}$ or $I={3\over 2}$.
The width of a $K^+\pi^-$ atom depends upon the difference between scattering
lengths in the two channels, $\Gamma\sim (a_{1/2}-a_{3/2})^2$,
(where $ a_{1/2}$ and $a_{3/2}$ are  the $I={1\over 2}$ and $I={3\over 2}$
scattering lengths, respectively)
while the shift of the ground-state depends upon a different combination, 
$\Delta E_0\sim 2 a_{1/2}+a_{3/2}$.

\begin{figure}
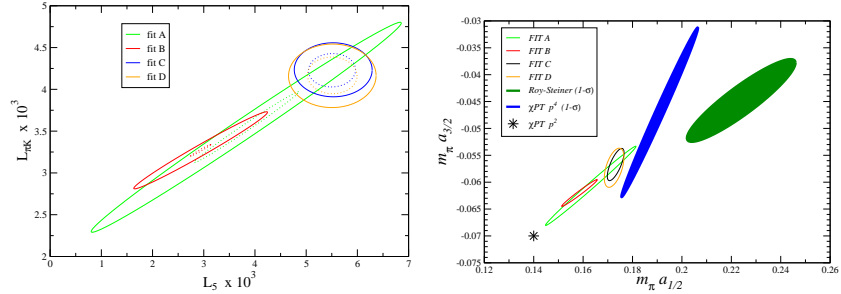

\vskip 0.2in
\begin{center}
\psfig{file=savage_1_c.eps,width=2.1in}\ \ \ \psfig{file=savage_1_d.eps,width=2.1in}
\end{center}
\caption{
The left panel shows the $95\% $ confidence ellipses for the counterterms $L_5$ and
  $L_{K\pi}$ that contribute to $K\pi$ scattering at NLO in $\chi$PT as
  extracted from the mixed-action lattice QCD calculation of NPLQCD at a
  lattice spacing of $b\sim 0.125~{\rm fm}$~\protect\cite{Beane:2006gj}.
The right panel shows the $95\% $ confidence ellipses for the $K\pi$ scattering
  lengths from a combination of a lattice QCD calculation and $\chi$PT
Also shown are the $38\%$ confidence ellipses from a Roy-Steiner 
analysis and from a ${\cal O}(p^4)$ $\chi$PT analysis.
}
\label{fig:LL}
\end{figure}
The NPLQCD collaboration calculated the $K^+\pi^+$ scattering length, $a_{K^+\pi^+}$,
in the
same way as we computed $\pi\pi$ scattering, requiring only the additional 
generation of strange quark valence propagators~\cite{Beane:2006gj}.
As $a_{K^+\pi^+}$ was calculated (again at a single lattice spacing of $b\sim
0.125~{\rm fm}$) at three different pion masses, a chiral extrapolation was
performed.  This extrapolation depends upon two counterterms, one from the
crossing-even and one from the crossing-odd amplitudes, and the important point
is that their coefficients have different dependence upon the meson masses.
Therefore, both can be determined from the results of the lattice calculation,
as shown in Fig.~\ref{fig:LL},
and therefore, the scattering lengths in {\bf both} isospin channels can be
predicted, as   shown in Fig.~\ref{fig:LL}.
This is a another demonstration of the combined power of lattice
QCD and $\chi$PT.

\subsection{Nucleon-Nucleon Scattering }

A few years ago we realized~\cite{Beane:2003yx,Beane:2003da} that even though
the scattering lengths in the
nucleon-nucleon system are unnaturally large, and much larger than the spatial
dimensions of currently available lattices, rigorous calculations in the
NN-sector could be performed today.
There are two aspects to this.  First, it is unlikely that the scattering
lengths in the NN sector are unnaturally large when computed on lattices with
the lightest pion masses that are presently available, 
$m_\pi^{\rm sea} \gsim 250~{\rm  MeV}$.
Second, it is not the scattering length that dictates the lattice volumes that
can be used in a Luscher-type analysis, but it is the range of the interaction,
which is set by $m_\pi$ for the NN interaction.
While the Luscher asymptotic formula are not applicable when the scattering
length becomes comparable to the spatial dimensions, the complete relation is
still applicable,
\begin{eqnarray}
p\cot\delta(p) \ =\ {1\over \pi L}\ {\bf S}\left(\,\frac{p L}{2\pi}\,\right)\ \
,\ \ 
{\bf S}\left(\,{\eta}\, \right)\ \equiv \ \sum_{\bf j}^{ |{\bf j}|<\Lambda}
{1\over |{\bf j}|^2-\eta^2}\ -\  {4 \pi \Lambda}
\ .
\label{eq:energies}
\end{eqnarray}
where $\delta(p)$ is the elastic-scattering phase shift, and
the sum in eq.~(\ref{eq:energies})
is over all triplets of integers ${\bf j}$ such that $|{\bf j}| < \Lambda$ and the
limit $\Lambda\rightarrow\infty$ is implicit~\cite{Beane:2003da}. 

The same set of configurations that was used to produce the scattering lengths
for $\pi\pi$ and $K\pi$ scattering, discussed above, were used to construct the
NN correlation functions, and hence to extract the energy shifts between two
nucleons at finite volume and twice the single nucleon mass at finite volume.
\begin{figure}
\vskip 0.2in
\begin{center}
\psfig{file=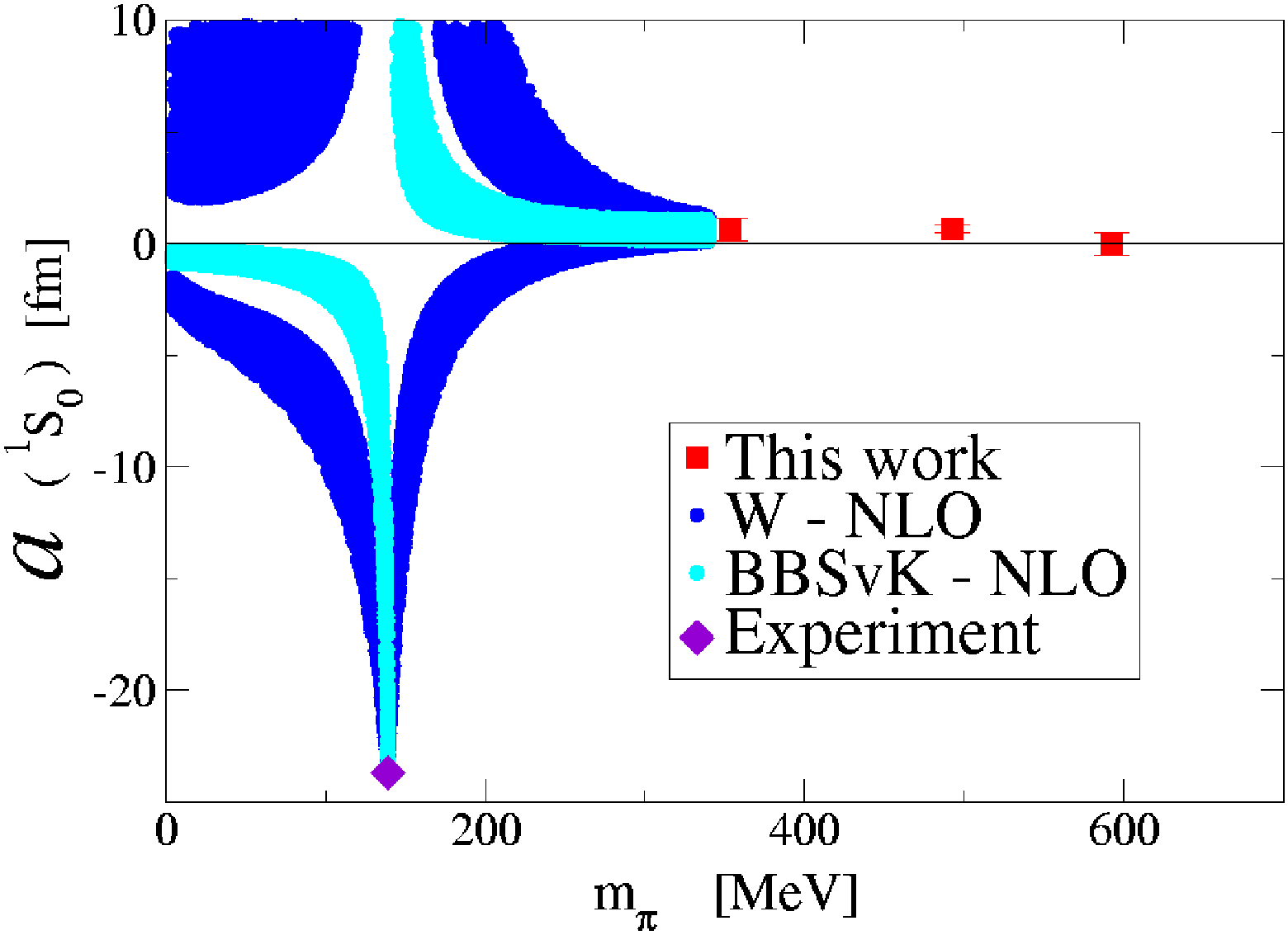,width=2.0in}\ \ \ \ \ 
\psfig{file=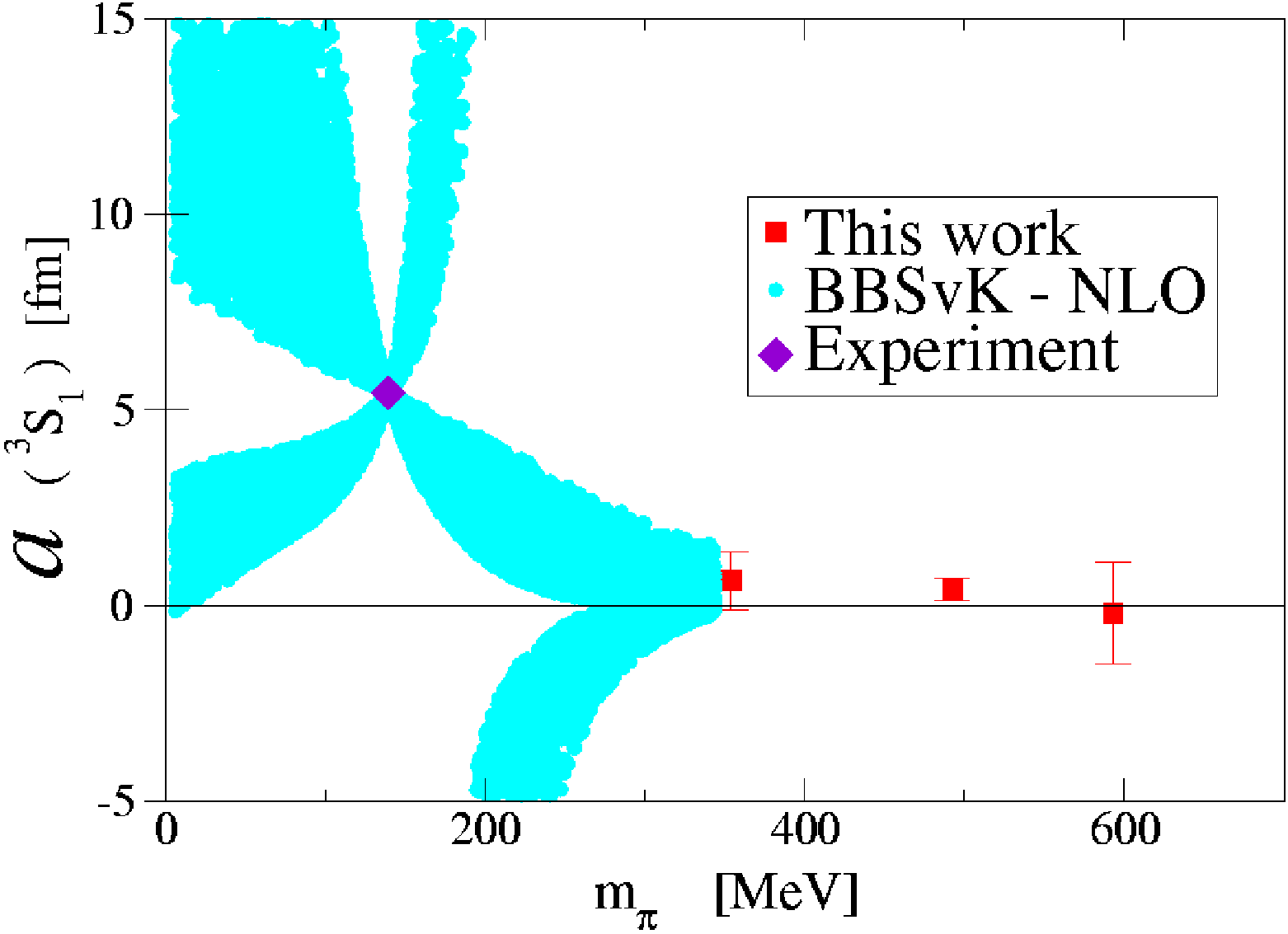,width=2.0in}
\end{center}
\caption{Scattering lengths in the $\si$ and $\siii-\diii$ NN channels a sa
  function of pion mass. The experimental value of the scattering length 
  and NDA have been used to constrain the  extrapolation in both 
BBSvK~\protect\cite{Kaplan:1998we,Kaplan:1998tg,Beane:2001bc} and 
W~\protect\cite{Weinberg:1990rz,Weinberg:1991um,Ordonez:1995rz} 
power-countings at NLO.
}
\label{fig:NN}
\end{figure}
At the pion masses accessible, the NN scattering lengths are found to be of natural size,
set by the inverse pion mass.
Only one of the pion masses is within the region described by the low-energy
EFT's, and as such a prediction of the scattering length from lattice QCD alone
cannot be made.  However, when combined with the physical values of the
scattering lengths, an allowed region for the scattering lengths as a function
of pion mass can be made, as shown in Fig.~\ref{fig:NN}.

\section{Hadronic Potentials}

Lattice QCD cannot, in general, isolate the potential between hadrons, as the
potential by itself is not an observable quantity.  An exception to this
statement is when both hadrons are infinite massive, and consequently their
spatial separation can be fixed (this of course generalizes to N-hadrons).
Currently, 
Silas Beane, William Detmold, Kostas Orginos and I are performing a quenched calculation
of the potential between two
B-mesons in the heavy quark limit.  
In this limit the light degrees of freedom (dof)
of each B-meson decouple from the heavy-quark dynamics, and their spin, $s_l$,  
becomes a good quantum number.  The potential between the two
B-mesons when constructed from $\chi$PT has the same form as that for NN, but
with different values of the counterterms that enter.
Therefore, calculating the potential between two B-mesons may provide
qualitative insight into the NN-system.
Our calculation is not the first attempt to extract the potential between heavy
hadrons, but is performed with a lattice spacing approximately half of that of
the previous calculations~\cite{Pennanen:1999xi,Cook:2002am}.
The quark mass was chosen to give $m_\pi\sim 420~{\rm MeV}$ and $m_\rho\sim
700~{\rm MeV}$.  
\begin{figure}
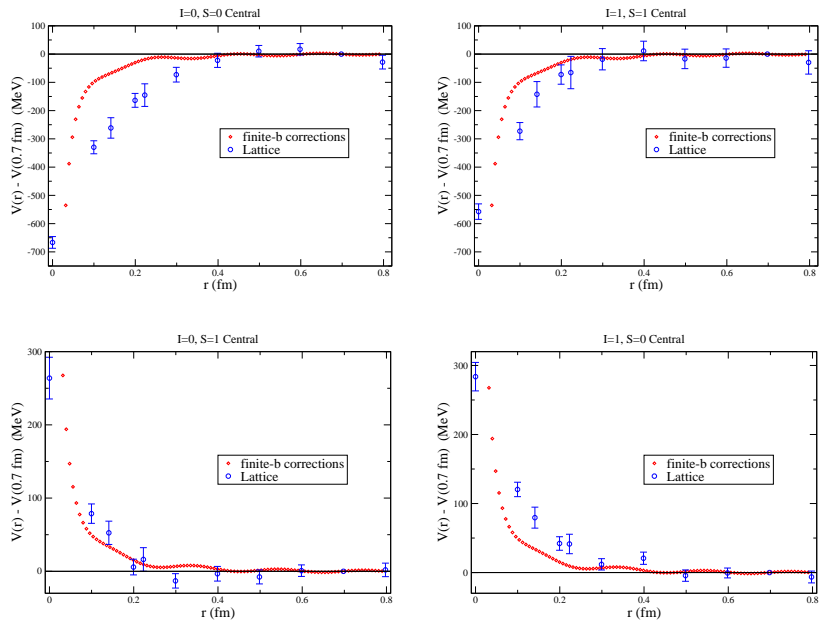

\vskip 0.2in
\begin{center}
\psfig{file=savage_1_g.eps,width=2.0in}\ \ \ \ \ 
\psfig{file=savage_1_h.eps,width=2.0in}\\
\vskip 0.2in
\psfig{file=savage_1_i.eps,width=2.0in}\ \ \ \ \ 
\psfig{file=savage_1_j.eps,width=2.0in}\\
\end{center}
\caption{
Preliminary calculations of the potential between two B-mesons in the heavy
quark limit. The vertical axis is the difference between the central potential
at a separation $r$ and that at $r=7 b$.  The fine (red) dots correspond to the
leading finite-lattice spacing correction that must be added to the potential
obtained in the lattice calculation.
}
\label{fig:BBcentral}
\end{figure}
Fig.~\ref{fig:BBcentral} shows our preliminary results for 
the central potentials as a function of B-meson
separation (the tensor potential is found to be very small, and so we have also
shown the potentials at $r=\sqrt{2}$ and $\sqrt{5}$ which are not  separated
into tensor and central). One clearly see short-distance repulsion in the channel with the
quantum numbers of s-wave NN interactions, while one finds large attraction in
the other channels.  There is a correction due to the finite lattice spacing
that is to be added to the results of the lattice calculation, and the leading
correction is
shown in Fig.~\ref{fig:BBcentral}.
Adding this factor gives the continuum potential between two infinitely massive B-mesons
in the  lattice volume, while  the extrapolation to infinite
volume potential has not yet been performed.

\begin{figure}
\vskip 0.1in
\begin{center}
\psfig{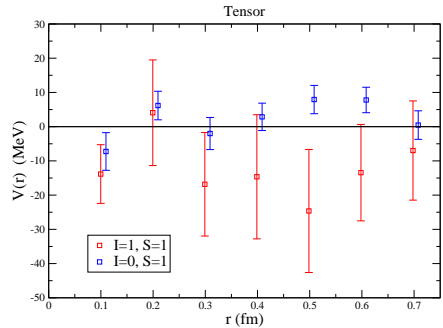}
\end{center}
\caption{
Preliminary calculations of the tensor potential between two B-mesons in the heavy
quark limit. 
}
\label{fig:BBtensor}
\end{figure}
The tensor interaction between hadrons has not been directly isolated in
previous works.  In Fig.~\ref{fig:BBtensor} we show our preliminary
calculation of the tensor potential between two infinitely massive B-mesons.
We find a potential that is consistent with zero within the uncertainty of the
calculation, indicating that the tensor potential in this system is no larger
than $V_T\lsim 40~{\rm MeV}$.

\section{The Three-Body Sector}

A discussion of the three-body sector would require multiple presentations all by itself.
I want to say just a  few words about it.
In order for lattice QCD to make connections with the three-body sector in
nuclear physics there must be both theoretical and numerical developments.
So far essentially none of the formalism has been developed with which to
analyze lattice results.  The analogous relations to the Luscher relation in the
NN sector must be developed, and numerically explored to determine parameters
that can be used in lattice calculations.  On the numerical side, it is clear
from the poor quality of the NN correlation functions obtained by NPLQCD, that
extracting any signal from the NNN sector will be very difficult, and likely
not possible on the coarse MILC lattices.  We are currently discussing
performing the contractions for the triton with the propagators we currently  have to quantify 
the situation.  Performing the contractions rapidly becomes computationally
expensive.  For the proton there are just 2 contractions, for pp there are 48
contractions while for the triton there are 2880 contractions, this is simply
$u! d!$.
At this stage we believe that the extraction scattering parameters from lattice QCD
in the three-body sector (in the near future)
will require the asymmetric Clover lattices that JLab have applied for computer resources to
produce during the next couple of years.

\section{Nuclear Effective Field Theory Lattice Calculations}

The last subject I wish to mention, but again I have no time to actually
discuss it,
is the recent development of performing EFT calculations on the lattice.
I think that this is an area that has a bright future and I refer those that
are interested to recent papers such as Refs.~\cite{Lee:2005it,Borasoy:2005yc}.

\section{Conclusions}

It is clear that lattice QCD is an important part of the future of nuclear
physics.
Lattice QCD, when combined with effective field theory,
is just starting to make rigorous predictions
of few-body observables, and we can expect significantly more progress as the
computational resources dedicated to these calculations is increased.
The formal tools are in place to explore the two-body sector, all that is
required is computational power.
However, a coherent effort involving both numerical calculations and formal developments
is presently required to move beyond the two-body sector.

\vskip 0.1in
I would like to thank my collaborators in this work, Silas Beane, Paulo
Bedaque, Kostas Orginos, William Detmold, 
Tom Luu, Elisabetta Pallante and Assumpta Parreno.

\end{document}